\documentclass[final]{svjour2}
\usepackage{graphicx}
\usepackage{rotating}
\usepackage{amssymb}
\usepackage{mathptmx}
\usepackage[numbers]{natbib}
\makeatletter
\journalname{Journal of Low Temperature Physics}
\begin{document}

\title{Experiments on the twisted vortex state in superfluid $^3$He-B}

\author{V.B. Eltsov \and R. de Graaf \and R.~H\"anninen \and M.~Krusius \and R.E.~Solntsev}

\institute{Low Temperature Laboratory, Helsinki University of Technology, P.O.Box
2200, 02015 HUT, Finland\\ Tel.: +358-9-4512973\\ Fax: +358-9-4512969\\
\email{ve@ltl.hut.fi}}

%\date{\today}
\date{14.07.2007}

\maketitle


\begin{abstract}

  We have performed measurements and numerical simulations on a
  bundle of vortex lines which is expanding along a rotating column of
  initially vortex-free $^3$He-B. Expanding vortices form a propagating front:
  Within the front the superfluid is involved in rotation and behind the front
  the twisted vortex state forms, which eventually relaxes to the equilibrium
  vortex state. We have measured the magnitude of the twist and its
  relaxation rate as function of temperature above $0.3\,T_{\rm c}$. We
  also demonstrate that the integrity of the propagating vortex front
  results from axial superfluid flow, induced by the twist.

\end{abstract}

\section{Introduction}

Since the pioneering works by Feynman, and by Hall and Vinen a
rotating superfluid has been associated with an array of rectilinear
vortex lines, stretched along the axis of rotation. This lowest energy
state is known as the equilibrium vortex state. Different kinds of
collective excitations of vortex arrays have been discussed over the
years \cite{sonin}, including analogs of Kelvin waves on bundles of
vortices \cite{vw,henderson} and Tkachenko waves
\cite{tkwpred,tkwobserv}. Recently a new state of vortex matter, the
twisted vortex state, was experimentally identified and theoretically
described in the B phase of superfluid $^3$He \cite{twistprl}. In this
state vortices have helical configuration circling around the axis of
rotation. Remarkably, certain such configurations, uniform in the
direction of the rotation axis, do not decay, although the energy of
this state is higher than that of the equilibrium vortex array. The
reason is that the force, acting on vortices, is directed along the
vortex cores everywhere.

\begin{figure}
\centerline{\includegraphics[width=\textwidth]{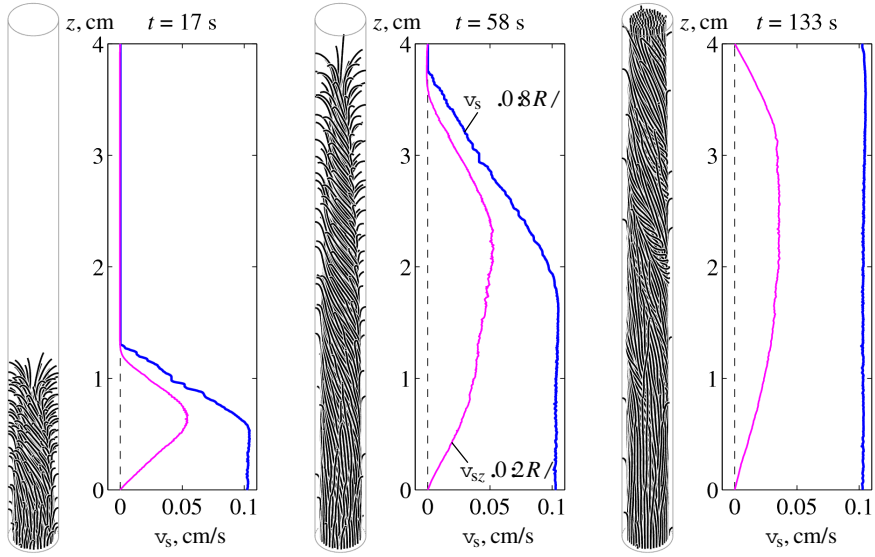}}
\caption{(Color online) Numerical simulations of vortex dynamics
  demonstrate the twisted vortex state during the expansion of the
  vortex bundle along the rotating column. The subsequent relaxation
  of the twisted state proceeds from the top and bottom sample
  boundaries. 
  The vortex configuration and the $z$-dependencies of the superfluid
  velocity components at given radial distance are shown at three
  points in time. The sample
  radius is 1.5\,mm, length 40\,mm, angular velocity 1\,rad/s and
  temperature $0.5\,T_{\rm c}$.}
\label{simul}
\end{figure}

The twisted vortex state is created when a bundle of vortex lines
expands along an initially vortex-free rotating superfluid column,
Fig.~\ref{simul}.  Those segments of the vortex lines, which terminate
on the side wall of the sample cylinder, propagate towards the
vortex-free part and simultaneously precess around the central axis
under the action of the Magnus and mutual friction forces. Such
two-component motion leads to a vortex bundle which is helically
twisted.  Expansion along the column becomes slower with decreasing
temperature as mutual friction decreases. Thus the spiral, created by
the motion of the vortex end on the side wall, becomes tighter. One
may expect that the resulting twist, characterized by the wave vector
$Q$ of the vortex helix in the bundle, becomes stronger as the
temperature is reduced. We have measured the magnitude of the twist in
the range 0.3\,--\,0.8\,$T_{\rm c}$ and observed this expected
behavior only at $T>0.45\,T_{\rm c}$. Below $0.45\,T_{\rm c}$ the
magnitude of the twist decreases again.

In the real sample, which is not infinitely long, the twist cannot be
completely uniform: At the top and bottom ends of the sample the
vortices are perpendicular to the wall and the twist disappears
there. The twist in the bulk unwinds when the vortex ends slide over
the end plates of the sample cylinder. The model in Ref.~\cite{twistprl}
predicts that the relaxation of the twist becomes faster with decreasing
temperature and distance to the wall. We have experimentally confirmed
both properties and established reasonable agreement with the model.

We also examine in this report the role of the twist-induced superflow
in the propagation of the vortex front, which separates the vortex-free
superfluid from the twisted vortex bundle. At $T\gtrsim 0.45\,T_{\rm
  c}$ the thickness of the vortex front increases while it
propagates. At lower temperatures the twist-induced axial superflow
pushes vortices at the rear of the front forward. Eventually they
catch up with the vortices in the head of the front and the front propagates
in a thin steady-state configuration.

\section{Numerical simulations}

The essential features of the vortex front and the twisted state can
be displayed by means of numerical calculations of vortex dynamics in
a rotating cylinder.  The simulation technique accounts fully for
inter-vortex interaction and for the effect of solid walls
\cite{simul}. In the initial state at $t=0$ the equilibrium number of
vortices is placed as quarter-loops between the bottom and the
cylindrical walls. During the subsequent evolution one observes the
formation of the vortex front and a twisted cluster behind it,
Fig.~\ref{simul}. The profile of the azimuthal component of the
superfluid velocity $v_{{\rm s}\phi}$ shows an almost linear
transition from the non-rotating state $v_{{\rm s}\phi} = 0$ to
equilibrium rotation $v_{{\rm s}\phi} \approx \Omega r$ within the
region of the vortex front. This shear flow within the
front is created by vortices which terminate on the cylindrical wall
perpendicular to the axis of rotation. At the temperature of
$0.5\,T_{\rm c}$ the thickness of the vortex front grows with
time. Simulations at $0.4\,T_{\rm c}$ on the other hand demonstrate
a thin time-invariant front \cite{twistprl}. We discuss this difference
below.

\begin{figure}
\centerline{\includegraphics[width=\textwidth]{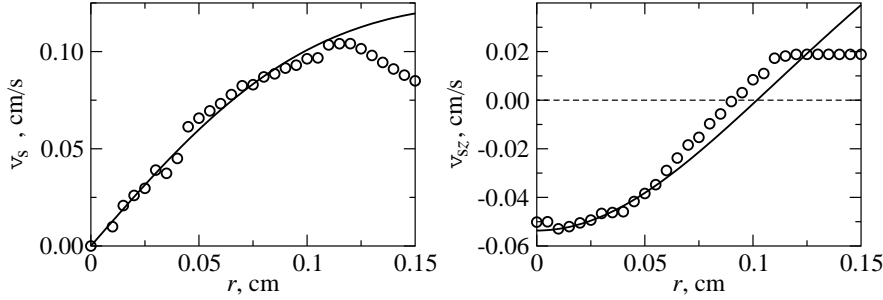}}
\caption{Radial dependencies of the superfluid velocity from the
  simulation snapshot in the center panel of Fig.~\ref{simul}. Circles
  are the simulation results at $z = 1.8\,$cm, averaged over the
  azimuthal angle. Lines are fit to the twist model Eq.~(\ref{uniftw})
  with $Q = 0.776 R^{-1}$.  At $r>0.12\,$cm the simulation results
  display the equilibrium vortex-free region around the central vortex
  cluster. This feature is ignored in the analytic model. }
\label{fitvs}
\end{figure}

The appearance of the twist is reflected in the axial superflow at the
velocity $v_{{\rm s}z}$, which is along the vortex expansion direction
close to the cylindrical boundary and in the opposite direction close
to the axis. As shown in Fig.~\ref{fitvs}, at constant height $z$ the
$r$-dependencies of $v_{{\rm s}\phi}$ and $v_{{\rm s}z}$ are
reasonably well described by the model suggested in
Ref.~\cite{twistprl}:
\begin{equation}
v_{{\rm s}\phi}(r)=\frac{(\Omega+Qv_0)r}{1+Q^2r^2},\ \
v_{{\rm s}z}(r)=\frac{v_0-Q\Omega r^2}{1+Q^2r^2},
\label{uniftw}
\end{equation}
where $v_0=(\Omega/Q)[Q^2R^2/\ln(1+Q^2R^2)-1]$ and $R$ is the sample
radius. The wave vector $Q$ of the twist has its maximum value close
to the rear end of the front and decreases to zero at the bottom and top
ends of the sample. 
The axial superflow affects the NMR spectrum of
$^3$He-B and this allows us to observe the twisted vortex state in the
experiment.

\section{Experiment}

\begin{figure}
\centerline{\includegraphics[height=0.4\textheight]{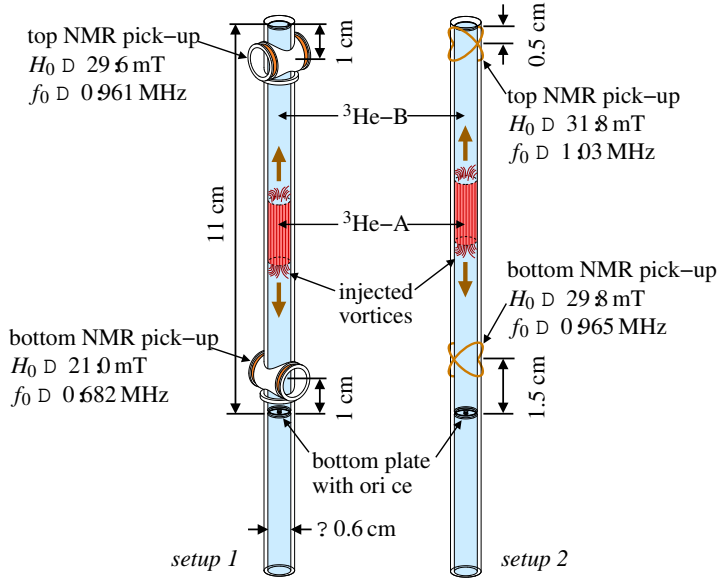}}
\caption{(Color online) Experimental setup.}
\label{expsetup}
\end{figure}

The experimental techniques are similar to those in the
recent studies of non-equilibrium vortex dynamics in $^3$He
\cite{turbreview}.  
The sample of $^3$He-B at 29\,bar pressure is
contained in a cylindrical smooth-walled container with dimensions
shown in Fig.~\ref{expsetup}. The
sample is split in two independent B-phase volumes by an
A phase layer, stabilized with applied magnetic field. In each B-phase
volume an independent NMR spectrometer is used to monitor the vortex
configuration. Two arrangements for pick-up coils have been used.
They are labelled throughout this report as ``setup 1'' and ``setup
2''.  These arrangements differ in the placement of the pick-up coils
with respect to the upper and lower ends of the container, in Larmor
frequencies, in the design of pick-up coils, and in the field
homogeneity. The last property is mostly determined by the field
distortion from the superconducting wire in the pick-up coils. For
solenoidal coils in setup 1 it is $\Delta H/H \approx 6\cdot10^{-4}$,
while the coils in setup 2 distort the field more and $\Delta H/H
\approx 1.7\cdot10^{-3}$.

The initial vortex-free state is prepared by thermal cycling of the
sample to temperatures above $0.7\,T_{\rm c}$, where one waits at
$\Omega = 0$ for the annihilation of the dynamic remanent vortices
left over from the previous experiment \cite{dyn-remn}. After that the
sample is cooled in rotation at $\Omega = 0.8\,$rad/s to the target
temperature.

\begin{figure}
\centerline{\includegraphics[width=0.9\textwidth]{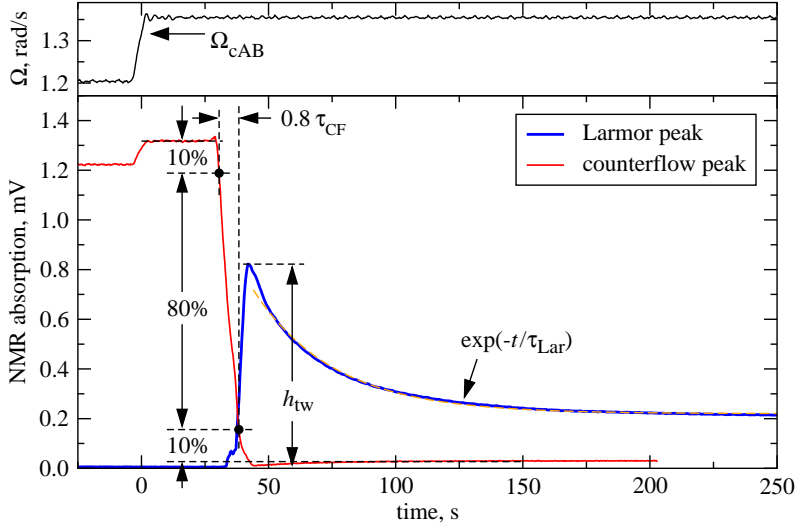}}
\caption{(Color online) Examples of NMR records from a vortex injection
  experiment and definition of the parameters $\tau_{\rm CF}$, $\tau_{\rm
    Lar}$ and $h_{\rm tw}$. The signal traces are recorded
  with the bottom spectrometer in setup 2 at $T=0.50\,T_{\rm c}$.}
\label{nmrtrace}
\end{figure}

For vortex injection the angular velocity is ramped with the rate
$\dot\Omega = 0.03\,$rad/s$^2$ above the critical velocity of the AB
interface instability $\Omega_{\rm cAB}$ \cite{ABinstab}, after which
$\Omega$ is kept constant, Fig.~\ref{nmrtrace}. In the instability
event about 10 vortices are injected into the B-phase close to the AB
interface. At $T<0.6\,T_{\rm c}$ a turbulent burst immediately follows
and generates almost the equilibrium number of vortices
\cite{measturb}. These vortices then propagate towards the pick-up coil. The
instability velocity $\Omega_{\rm cAB}$ is controlled by temperature
and magnetic field profile and varies between 1.1 and 1.5\,rad/s. A
modification of this injection technique has been used at
$T>0.6\,T_{\rm c}$. Here the sample initially rotates at $\Omega >
2\,$rad/s without the A phase. Then the magnetic field is increased
until the formation of the A phase starts. The AB interface
immediately goes unstable and hundreds of vortices are generated
\cite{injLT}.

While vortices expand along the sample column and pass through the
pick-up coil their configuration is read using NMR. The signal is
measured at a fixed frequency either at the location of the
counterflow or the Larmor peak in the NMR spectrum
\cite{measturb}. Absorption at the counterflow peak is sensitive to
the difference in azimuthal flow velocities of the normal and
superfluid components $v_{{\rm n}\phi}-v_{{\rm s}\phi}$. It is at
maximum in the initial vortex-free state with $v_{{\rm
    s}\phi}=0$. When $v_{{\rm s}\phi}$ increases the absorption
decreases rapidly and drops to zero when $v_{{\rm s}\phi} \sim (1/2)
v_{{\rm n}\phi}$ \cite{juhacalc}.  Absorption in the Larmor peak is
mostly sensitive to the axial flow $v_{{\rm s}z}$. It is zero in the
vortex-free state while for a vortex cluster it has some finite value,
which increases monotonously with increasing $v_{{\rm s}z}$
\cite{juhacalc}.

An example of NMR measurement is presented in Fig.~\ref{nmrtrace}. To
record signals at both Larmor and counterflow peaks with the same
spectrometer the experiment is repeated twice in identical
conditions. A rapid drop in the counterflow signal is seen which
corresponds to the passage of the vortex front through the coil. We
characterize this drop with time $\tau_{\rm CF}$. The Larmor peak
grows first to a maximum value $h_{\rm tw}$, which corresponds to the
maximum twist. Then the signal relaxes exponentially with the time
constant $\tau_{\rm Lar}$ to a value characteristic for an equilibrium
vortex cluster. The propagation velocity $V_{\rm f}$ of the vortices
at the head of the front can be determined from their flight time
between the injection moment, where $\Omega(t) = \Omega_{\rm cAB}$,
and the arrival to the edge of the peak-up coil, as monitored with the
counterflow peak. We do not discuss $V_{\rm f}$ in detail in this
report.

\section{Results}

\begin{figure}
\centerline{\includegraphics[width=0.75\textwidth]{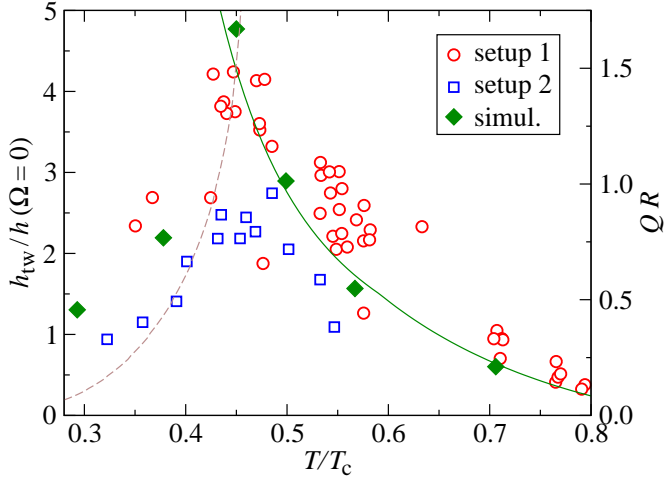}}
\caption{(Color online) Strength of the twist as a function of
  temperature.  The measurements are performed using the bottom
  spectrometer in both setups of Fig.~\ref{expsetup}. The ratio of the
  maximum amplitude of the Larmor peak in the twisted state to the
  amplitude of the Larmor peak in the nonrotating sample is plotted
  on the left axis. The maximum value of the twist wave vector $Q$,
  obtained in simulations, is plotted on the right axis. The solid curve
  shows the fit $Q\, R = 0.7 (1-\alpha')/\alpha$. The broken curve shows the
  minimum magnitude of the twist for which propagation of the vortex
  front in a thin steady-state configuration is possible.}
\label{overshoot}
\end{figure}

The temperature dependence of the magnitude of the twist is presented
in Fig.~\ref{overshoot}. From the experiment the raw data are plotted:
The maximum height of the Larmor peak $h_{\rm tw}$
(Fig.~\ref{nmrtrace}), normalized to the height of the Larmor peak at
$\Omega=0$ at the same temperature. The data obtained in setup~1 and
setup~2 are not identical since the spectrum shape around the Larmor
peak depends on the magnitude of the magnetic field via the order
parameter texture and also the homogeneity of the magnetic field is
important: In setup 2 the homogeneity is worse and the Larmor peak
cannot grow as high (i.e. as narrow) as in setup 1.  Anyway, a clear
maximum in the twist-induced signal is observed at around $0.45T_{\rm
  c}$.  The same behavior of the magnitude of the twist is confirmed
in the numerical simulations: The maximum value of the twist wave vector
behind the front, as determined from a fit of the velocity profiles to
Eq.~(\ref{uniftw}), also peaks at $0.45T_{\rm c}$.

The initial growth of the twist with decreasing temperature is
expected. From the equations of motion of a single vortex which ends on the
cylindrical wall, the estimate for the expansion velocity along
$z$ is 
\begin{equation}
  v_{{\rm L}z} = \alpha(T)\,[v_{{\rm n}\phi}(R) - v_{{\rm s}\phi}(R)],
\label{vLz}
\end{equation}
where $\alpha$ is a mutual friction coefficient \cite{bevan}. As the
normal fluid velocity at the side wall $v_{{\rm n}\phi}(R) = \Omega R$
and the superfluid velocity induced by a single vortex can be
neglected in our conditions, Eq.~(\ref{vLz}) gives $v_{{\rm L}z}
\approx \alpha \Omega R$. For the azimuthal velocity of the vortex end
in the rotating frame a similar estimation gives $v_{{\rm L}\phi}
\approx (1-\alpha') \Omega R$, where $\alpha'$ is another mutual
friction coefficient \cite{bevan}. Thus the trajectory of the vortex end on the
side wall is a spiral with wave vector $Q = v_{{\rm L}\phi}/(R
v_{{\rm L}z}) \approx (1-\alpha')/(R \alpha)$. This value can be used
as a rough estimation of the $Q$ vector in the twisted vortex
state. As one sees from Fig.~\ref{overshoot}, the simulation results at $T
\gtrsim 0.45\,T_{\rm c}$ indeed follow this dependence.

For the decreasing twist at $T < 0.45\,T_{\rm c}$ a couple of reasons
can be suggested. First, the twist can relax through reconnections
between vortices in the bundle, which become more prominent with
decreasing temperature. Second, relaxation of the twist proceeds in
diffusive manner within the twisted cluster. The source of the twist
is at the vortex front, while the sink is at the end plate of the
cylinder, where the twist vanishes because of the boundary
conditions. The effective diffusion coefficient increases
as the temperature decreases \cite{twistprl}. The faster diffusion limits
the maximum twist in the finite-size sample at low
temperatures.

\begin{figure}
\centerline{\includegraphics[width=0.65\textwidth]{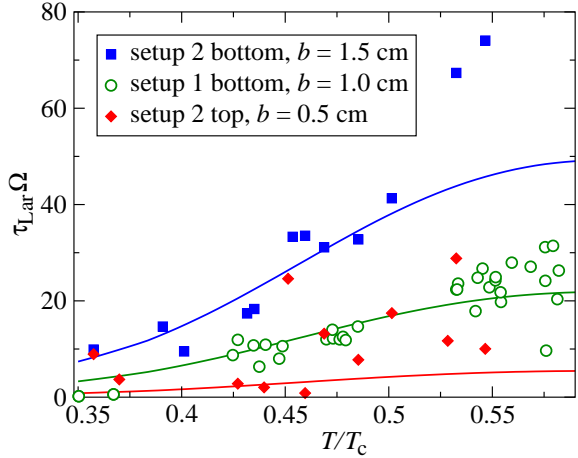}}
\caption{(Color online) Relaxation time of the twist versus
  temperature. Solid lines are fit to Eq.~(\ref{relaxtau}) with $C=2$.}
\label{relax_time}
\end{figure}

The relaxation of the twisted state is observed as a decay in the
amplitude of the Larmor peak. The measured time constant $\tau_{\rm
  Lar}$ of the exponential decay is presented in Fig.~\ref{relax_time}
for three different positions of the detector coil with respect to the
end plate of the sample cylinder. It is clear that the relaxation
indeed becomes faster with decreasing temperature. It is also evident
that the relaxation of the twist proceeds faster as the observation
point moves closer to the end plate of the sample cylinder. Both of
these features are well described by the model presented in
Ref.~\cite{twistprl}. According to this model the relaxation time of
the twist is
\begin{equation}
\tau \approx \frac{C}{\Omega} \left( \frac{b}{R} \right)^2
\frac{\alpha}{[(1-\alpha')^2+\alpha^2]},
\label{relaxtau}
\end{equation}
where $C\sim1$ and $b$ is the distance along $z$ to the end plate,
where the twist vanishes. The fit to this model using $C$ as the only
fitting parameter demonstrates reasonable agreement with the
measurements in Fig.~\ref{relax_time}. 
The experimental points show
somewhat faster temperature dependence than the model. The discrepancy
may result from shortcomings of both
the model and the experiment. The model is constructed for the case of
weak twist ($QR\ll1$). In the experiment the probe is not point-like,
but has a height comparable to the value of $b$ itself.

When Eq.~(\ref{vLz}) is applied to vortices within the vortex front,
the following problem arises: At the head of the front the
superfluid component is almost at rest, $v_{{\rm s}\phi} \ll v_{{\rm
    n}\phi}$, and the expansion velocity of vortices is
$V_{\rm f} \approx \alpha \Omega R$. Behind the front the density of
vortices is close to the equilibrium and $v_{{\rm s}\phi} \approx
v_{{\rm n}\phi}$. Thus vortices at the tail of the front are barely
able to
expand and the thickness of the front should
rapidly increase in time. On the other hand simulations show that
vortices at the tail of the front do expand,
Fig.~\ref{simul}. Moreover, at sufficiently low temperatures the
expansion velocity of these vortices reaches the velocity of the
foremost vortices, so that the front propagates in a thin steady-state
configuration \cite{twistprl}.

The explanation is that the vortex state behind the front is not the
equilibrium state, but the twisted vortex state. Taking into account
the axial superflow, induced by the twist, Eq.~(\ref{vLz}) should be
modified as $v_{{\rm L}z} = \alpha\,[v_{{\rm n}\phi}(R) - v_{{\rm
    s}\phi}(R)] + (1-\alpha')v_{{\rm s}z}(R)$. Given that $v_{{\rm
    s}z}(R)$ is in the direction of the front propagation and $v_{{\rm
    s}\phi}(R) <v_{{\rm n}\phi}(R)$ in the twisted state, the
expansion velocity $V_{\rm t}$ of the vortices in the tail of the front is
enhanced. This velocity can be estimated taking $v_{{\rm s}z}(R)$ and
$v_{{\rm s}\phi}(R)$ from Eq.~(\ref{uniftw}):
\begin{equation}
V_{\rm t} = \alpha \Omega R 
\left[ 1 + \frac{1-\alpha'}{\alpha}\,\frac{1}{Q\,R} \right]\,
\left[ 1 - \frac{Q^2R^2}{(1 + Q^2R^2)\,
\log(1 + Q^2R^2)} \right]
\label{Vt}
\end{equation}
This velocity has a maximum as a function of the $Q$ vector. If
$\alpha/(1-\alpha') > 0.52$ (i.e. $T>0.46\,T_{\rm c}$ \cite{bevan}),
the maximum value of $V_{\rm t}$ is less than the velocity of the
foremost vortices $V_{\rm f} \approx \alpha \Omega R$. In these
conditions the thickness of the front increases while it
propagates. When $T<0.46\,T_{\rm c}$ a wide range of $Q$ values exists
for which formally $V_{\rm t} \geqslant V_{\rm f}$. The minimum
possible value of $Q$ is shown in Fig.~\ref{overshoot} by the broken
curve. In these conditions the vortex front propagates in a
steady-state ``thin'' configuration. These simple considerations
become inapplicable, however, when
vortices interact strongly within the thin layer. To determine, say,
the stable front thickness and the magnitude of the twist in
the regime below $0.46\,T_{\rm c}$ a different approach would be needed.

The change in the front propagation at $T\approx 0.45\,T_{\rm c}$ is
observed in both experiment and simulations. In the experiment the
decay time of the counterflow peak $\tau_{\rm CF}$ in
Fig.~\ref{nmrtrace} can be used to extract the front thickness. The
decay of the counterflow peak starts when the head of the vortex front
arrives at that edge of the detector coil, which is closer to the
injection point. The counterflow signal vanishes when the last part of
the front, which still possesses enough counterflow to generate the NMR
response, leaves the far edge of the detector coil. The product
$\tau_{\rm CF} V_{\rm f}$ has the dimension of length and can be
called the apparent thickness of the front.  When the real thickness
of the front grows with time its apparent thickness depends on the
distance of the observation point from the injection point and on the
rate with which the real thickness increases. When the front remains
thin the value of its apparent thickness equals the height of the
pick-up coil $d_{\rm coil} = 9\,$mm.

\begin{figure}
\centerline{\includegraphics[width=0.65\textwidth]{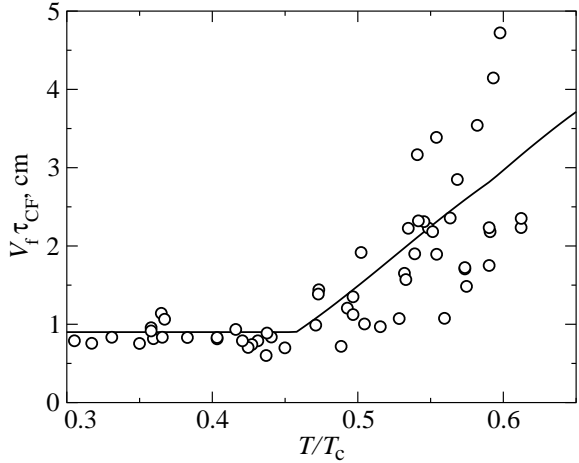}}
\caption{Apparent thickness of the vortex front as function of
  temperature. The solid line is the prediction of the model in
  Eq.~(\ref{thickmodel}).}
\label{front_thick}
\end{figure}

The measurements of the apparent thickness of the front are presented
in Fig.~\ref{front_thick}. The scatter at higher temperatures has two
sources. Partially it is due to the uncertainty in the determination
of $V_{\rm f}$ from flight times which are not long compared to the
measuring resolution. Another contribution is the variation in the
decay profile of the counterflow signal which may display weak
oscillations around a roughly linear decrease. At $T>0.45T_{\rm c}$ we
have $\tau_{\rm CF} V_{\rm f} > d_{\rm coil}$ and the apparent
thickness decreases with temperature. Finally at $T < 0.45\,T_{\rm
  c}$ the front becomes thin compared to the height of the pick-up
coil.  Assuming that at the moment of injection the front is
infinitely thin we can write
\begin{equation}
\tau_{\rm CF} = \frac{d + d_{\rm coil}}{V_{\rm t}^*} - \frac{d}{V_{\rm f}},
\label{thickmodel}
\end{equation}
where $d$ is the distance from the injection point (i.e. position of the AB
interface) to the nearest edge of the pick-up coil and $V_{\rm t}^*$ is the
expansion velocity at the position in the front where the NMR signal from the
counterflow vanishes. Given that the latter condition roughly
corresponds to $v_{{\rm s}\phi} \sim (1/2) v_{{\rm n}\phi}$ we take $V_{\rm
  t}^* = (V_{\rm t}+V_{\rm f})/2$ if $V_{\rm t} < V_{\rm f}$ and simply
$V_{\rm t}^* = V_{\rm f}$ otherwise. Using $V_{\rm t}$ from Eq.~(\ref{Vt})
and the simple estimates $QR = (1-\alpha')/\alpha$ and $V_{\rm f} = \alpha
\Omega R$, we get from Eq.~(\ref{thickmodel}) the solid line in
Fig.~\ref{front_thick}, which is in reasonable agreement with the
experiment.

In the simulations the thickness of the front grows with time at
higher temperatures. This process slows down as the twist increases with
decreasing temperature (Fig.~\ref{overshoot}). Finally at $T\approx
0.45\,T_{\rm c}$ the twist reaches the value which is
enough, according to Eq.~(\ref{Vt}), to support a thin front
configuration. At lower temperatures the thickness of
the front indeed becomes time-independent and roughly equal to the
radius of the sample. Simultaneously the twist behind the
front starts to drop, as has been discussed above. It remains,
however, within the limits where the twist-induced superflow is able
to keep the front thin.

\section{Conclusions}

We have studied the formation and relaxation of the twisted vortex
state in $^3$He-B in the temperature range between $0.3\,T_{\rm c}$
and $0.8\,T_{\rm c}$. At higher temperatures $T\gtrsim 0.45\,T_{\rm
  c}$ the twist behind the propagating vortex front grows with
decreasing temperature as $Q \propto (1-\alpha')/\alpha$. Here the
thickness of the front increases while it propagates along the
rotating column. The axial superflow, induced by the twisted state,
boosts the expansion velocity of vortices in the tail of the
front. This enhancement increases as the twist grows with
decreasing temperature. Finally at $T\approx 0.45\,T_{\rm c}$ vortices
in the tail of the front are able to catch up with vortices in the head. At
lower temperatures the front propagates in a thin steady-state
configuration, while the twist starts to decrease.

The relaxation of the twist in a sample of finite length is not
related to the front. It proceeds from the walls which limit the
length of the sample along the rotation axis. The relaxation
speeds up with decreasing temperature, unlike many other processes in
vortex dynamics.

While the understanding of the front
propagation and of the formation of the twisted vortex state in the
high-temperature regime $T\gtrsim0.45\,T_{\rm c}$ is quite good, in the
low-temperature regime of a thin front the theoretical
understanding is lacking. Especially interesting would be to consider
the role of vortex reconnections and turbulence, both of which should increase
with decreasing temperature.

\end{document}